\documentclass[conference]{IEEEtran}
\makeatletter
\def\ps@headings{%
\def\@oddhead{\mbox{}\scriptsize\rightmark \hfil \thepage}%
\def\@evenhead{\scriptsize\thepage \hfil \leftmark\mbox{}}%
\def\@oddfoot{}%
\def\@evenfoot{}}
\makeatother
\usepackage{booktabs} % For formal tables
\usepackage{url}
\usepackage{svg}
\usepackage{etoolbox}
\usepackage{cuted}
\usepackage{lipsum}
\usepackage{array}
\usepackage{amsmath}
\usepackage{tikz}
\usepackage{color}
\usepackage{dirtytalk}
\usepackage{algpseudocode}
\usepackage{algorithm}
\usepackage{graphicx}
\ifCLASSINFOpdf
\else
\fi
\begin{document}
%\title{DDRAV: Decentralized Data Recorder based on Proof of Location and Authority with Union Consensus for Autonomous Vehicle}
\title{Blockchain-inspired Event Recording System \\ for Autonomous Vehicles}

% Blockchain-inspired Event Recorder for Autonomous Systems

%Decentralized Data Recorder based on Proof of Event and Dynamic Federation Consensus
\author{\IEEEauthorblockN{Hao Guo ~~~~ Ehsan Meamari~~~~ Chien-Chung Shen}
\IEEEauthorblockA{Department of Computer and Information Sciences\\
University of Delaware, U.S.A. \\
% Newark, DE 19716\\
\{haoguo,ehsan,cshen\}@udel.edu}}

%Telephone: (800) 555--1212, Fax: (888) 555--1212}


\maketitle

\begin{abstract}
Autonomous vehicles are capable of sensing their environment and navigating without any human inputs. 
% However, with autonomy, comes accountability. 
However, when autonomous vehicles are involved in accidents between themselves or with human subjects, 
liability must be indubitably decided based on accident forensics.
This paper proposes a blockchain-inspired event recording system for autonomous vehicles. Due to the inefficiency and limited usage of certain blockchain features designed for the traditional cryptocurrency applications, we design a new ``proof of event'' mechanism to achieve indisputable accident forensics by ensuring that event information is trustable and verifiable. Specifically, we propose a dynamic federation consensus scheme to verify and confirm the new block of event data in an efficient way without any central authority. The security capability of the proposed scheme is also analyzed against different threat and attack models.
\end{abstract}


% \fi
\IEEEpeerreviewmaketitle

\section{Introduction}
% The autonomous system performs tasks with a high degree of autonomy. Our proposed system is proper for every autonomous system but, in this paper, we use autonomous vehicle as the platform to illustrate the new idea. 
An autonomous vehicle (also known as driverless vehicle or self-driving vehicle) is capable of sensing its environment and navigating without any human input \cite{gehrig1999dead}.
To facilitate self-driving, autonomous vehicles adopt a variety of sensory technologies, such as lidar, camera, and ultrasound, to detect their surroundings, and use a control system to interpret sensory information to identify appropriate navigation paths, as well as avoid obstacles and follow relevant traffic signs.
Based on a recent survey, roughly two-thirds of Americans expect cars to be totally autonomous in the next half century~\cite{survey}. 

However, {\em with autonomy, comes accountability}. When autonomous vehicles are involved in accidents (collisions between themselves, or collisions
with conventional vehicles, pedestrians or other objects), how could such events be recorded for forensic purposes to determine liability? In addition, how could such recorded events be verified, trusted, and not tampered? Such issues become critical when there exist incentives for different parties involved to tamper with the recorded events to avoid punitive penalties. This paper proposes a blockchain-inspired event recording scheme to help autonomous vehicles achieve a tamper-proof and verifiable event recording and forensics system.

A blockchain consists of a series of blocks, each of which is composed of sets of timestamped transactions and a hash of its previous blocks~\cite{dilley2016strong}. 
The original blockchain was designed for Bitcoin, a digital cryptocurrency, to solve the double-spending problem by using the {\it Proof of Work} (PoW) mechanism~\cite{tschorsch2016bitcoin}.  In PoW, miners compete with one another to become the first to solve a hash puzzle so as to obtain the right to generate the next block and to receive incentives. However, PoW usually takes 10 minutes to solve a puzzle and generate a new block. Due to the computational difficulty of the current PoW, miners tend to form bigger mining pools to conduct PoW~\cite{eyal2015miner}, which diminishes one original feature of being decentralized. 

In our proposed event recording system, accidents are recorded as timestamped transactions which are to be saved into a new block in real-time.
Although autonomous vehicles may be equipped with reasonable computing capacity, conducting PoW to save the event of an accident in real-time will not be feasible due to the complexity of solving a hash puzzle. 

To address this critical issue, we propose the mechanism of {\em Proof of Event with Dynamic Federation Consensus} to record accident events in a new block. When an accident occurs, vehicles directly involved in the accident broadcast `event generation' requests (via IEEE 802.11p [DSRC], for instance), which only those vehicles within the (DSRC) communication range will receive and respond. Then, both the vehicles directly involved in the accident and those vehicles receiving the request will generate and broadcast the event into a `vehicular network' which is defined based on the existing cellular network infrastructure. Within the vehicular network, a random federation group is formed to verify and save the event data into a new block by using a multi-signature scheme~\cite{bitcoinmulti}. And finally the generated new block will be sent and saved in Department of Motor Vehicles (DMV) for the permanent records. 

% In summary, contributions of the paper are as follows:

The mechanism of Proof of Event with Dynamic Federation Consensus records events for indisputable accident forensics and protects data integrity and trustworthiness by utilizing event data from multiple sources and the generated hash digest. The recorded events also provide traceable evidence.
%with the help of multiple autonomous vehicles and the generated hash digest for event information.
Specifically, the proposed Dynamic Federation Consensus scheme replaces the role of PoW in the original blockchain to confirm and save a new block in a fast and effective way without incurring extensive computation. As a federation is dynamically formed around each accident over a vehicular network, the consensus on the authenticity of the generated events can be recorded in a flexible and robust manner.

The remainder of the paper is organized as follows. Related work is described in Section \ref{sec:related}. In Section \ref{sec:model}, we present the cellular network-based vehicular network and describe the mechanism of Proof of Event  with Dynamic Federation Consensus. In addition to normal cases, `extreme' accident scenarios are discussed in Section \ref{sec:extreme}. In Section \ref{sec:security}, we analyze the security of the proposed scheme against potential attacks. Section \ref{sec:conclusion}  concludes the paper.

\section{Related Work}
\label{sec:related}

\subsection{Event Data Recorders}

An event data recorder (EDR), a vehicle equivalent of a plane's flight recorder or ``black box,'' is installed in vehicles to record information related to crashes or accidents~\cite{wiki:edr}. Some EDRs continuously record data until a crash or accident stops them, and others are activated by crash-like events (such as a sudden decrease in velocity) and may continue to record until the accident is over, or until the recording time expires~\cite{wiki:edr}. 
%To the best of our knowledge, there is no existing EDR integrates the blockchain-inspired technology to achieve the immutability and decentralized trustworthy.
%The Bosch EDR tool  allows re-image crash data directly from all supported vehicles giving a detailed report of critical data parameters leading up to and during a crash~\cite{wiki:traffic}.
Due to its individual and independent installation, once an EDR is damaged or malfunctions, there is no chance to restore or verify the information stored.
%and if the EDR manipulates the data log there is no protection and detection mechanism. 

Heijden et al.~\cite{DBLP:journals/corr/abs-1710-08891} proposed a distributed ledger that provides accountability for both misbehavior authorities and vehicles. The goal is to reduce the requirements of trust on users of vehicular communication systems and to create accountability for misbehavior authorities via hierarchical consensus and global revocation.
% blockchain-based distributed EDR system to achieve message and revocation accountability by forming a compressed global state~\cite{DBLP:journals/corr/abs-1710-08891}. 
In contrast, our work focuses on accident forensics for autonomous vehicles. By employing the mechanism of Proof of Event with Dynamic Federation Consensus, accident events are stored in a trustable, verifiable, and tamper-proof manner.

\subsection{Blockchain-based Vehicular Systems}

Yuan et al.~\cite{yuan2016towards} proposed a secured and decentralized blockchain-based autonomous intelligent transportation systems with better usage of infrastructures and resources. A case study is presented to describe a blockchain-based real-time
ride-sharing system. By using Ethereum’s smart contract system. Leiding et al.~\cite{leiding2016self} proposed a self-managed and decentralized system to deploy and run any type of application on vehicular ad-hoc networks without a central managing authority. 
By using a blockchain-based public key infrastructure,
Rowan et al.~\cite{rowan2017securing} proposed a novel inter-vehicle session key establishment protocol to secure vehicle-to-vehicle communications through visible light and acoustic side-channels. 
However, none of the work took inspiration from blockchain to design an accident forensics system for autonomous vehicles.

\subsection{Consensus Methods}

As consensus is critical to the decentralized nature of blocchain, we review existing consensus schemes to highlight our unique contribution. In the current state-of-the-art, PoW~\cite{Jakobssonpow}, Proof of Stake (PoS)~\cite{vasin2014blackcoin}, and Proof of Authority (PoA)~\cite{de2017pbft} and several other Proof of `X' consensus models all rely on selecting one single peer to produce the new block. However, these consensus models gradually deviate from the original goals of decentralization and democratization. For instance, one single peer is selected by “nonce lottery” via mining as with PoW, by random selection among the largest stake holders as with PoS~\cite{ucoin}, and by random selection among nodes via one centralized authority as with PoA. Therefore, large mining pools centralize authorities of Bitcoin, PoS concentrates power in the hands of few peers based on their balance, and PoA leaves the decision of which entities can generate new blocks in the network to one central authority~\cite{ucoin}. Thus, all of these models have been falling short of their initial goals.
In contrast, the mechanism of Proof of Event with Dynamic Federation Consensus proposed in this paper addresses the dynamic and autonomous nature of self-driving vehicles so that the accident forensic information could be validated by a federation than one single individual.  

\section{Architecture of Recording System}
\label{sec:model}

\subsection{Cellular Network-based Vehicular Network}
\label{sec:cell}

\begin{figure}[h]
\includegraphics[width=0.48\textwidth]{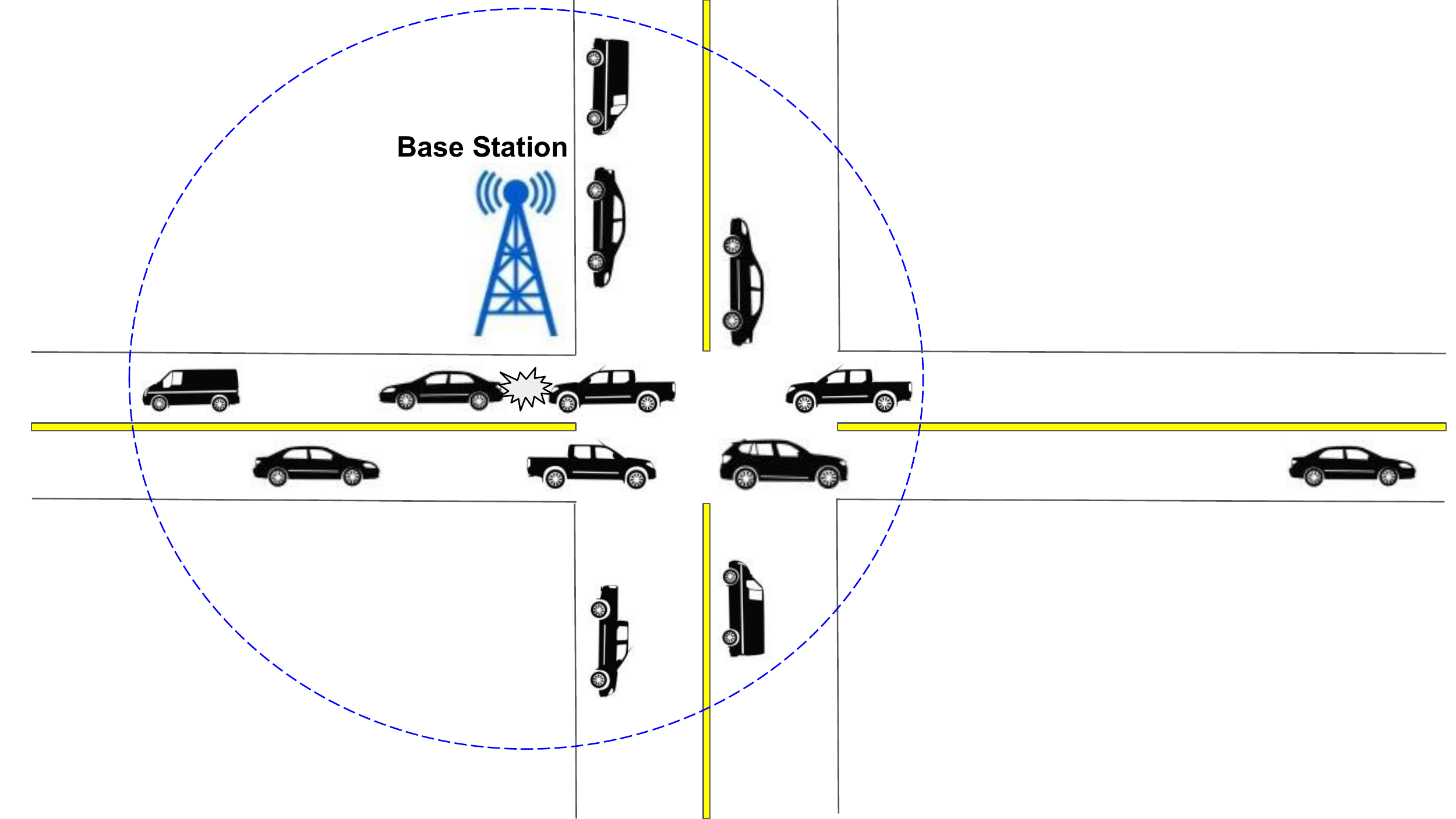}
\centering
\caption{Cellular network-based `vehicular' network.}
\label{fig:1}
\end{figure}

We adopt a cellular network-based infrastructure to define a `vehicular network' for each accident, where all the vehicles covered by the same base station ({\em i.e.}, within the same cell) of the vehicles directly involved in an accident form a `vehicle network' as depicted in Fig. \ref{fig:1}. We also assume that all the autonomous vehicles  register with an authority, such as Department of Motor Vehicles (DMV), using license plate and VIN number. 

Vehicles use the IEEE 802.11p standard of Dedicated Short-Range Communications (DSRC)~\cite{jiang2006design} to send and receive `event generation' requests. Meanwhile, vehicles are connected to a cellular network to broadcast and confirm event data within the corresponding vehicular network
so as to create new blocks.

%By adopting this idea, we have the dynamic vehicular network structure based on the real geographic location of neighboring cars.
%\textbf{(I think that "neighboring" and "overlapping" are two similar expressions. Just use one of them for whole of paper.)}
% * <ehsan@udel.edu> 2018-03-29T01:52:21.465Z:
% 
% I think we should merge section 3 and 4, because section 3 is not big enough to be a section.)
% 
% ^ <ehsan@udel.edu> 2018-04-07T09:16:13.142Z.

\begin{figure*}[bt]
\centering
\includegraphics[width=0.9\textwidth]{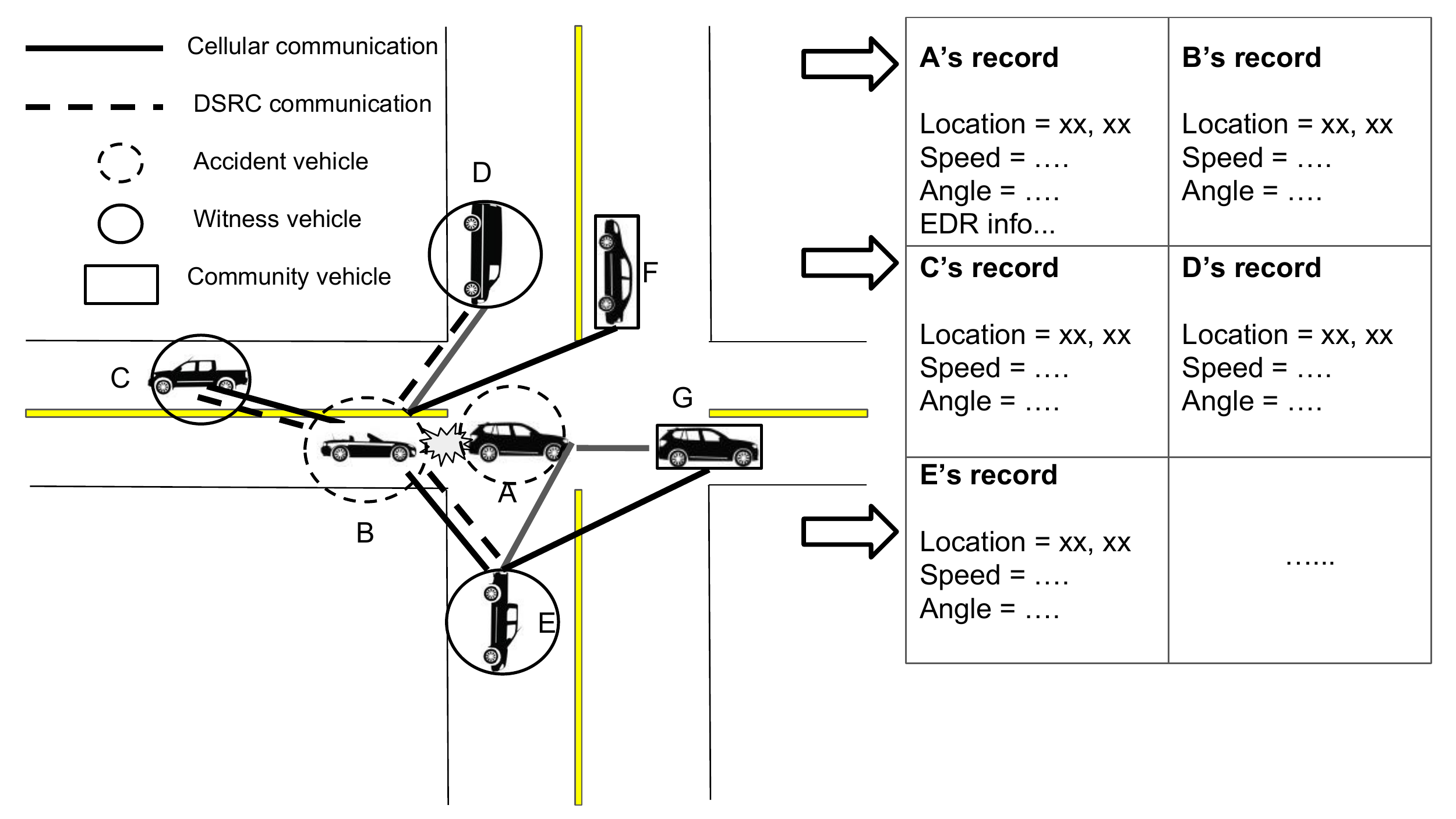}
\centering
\caption{\label{fig:accident}After an accident, both accident and witness vehicles generate and broadcast event data.}
\end{figure*}

\subsection{Proof of Event with Dynamic Federation Consensus}

To facilitate forensic investigation after an accident, one issue is the correctness and trustworthiness of the recorded data, as vehicles involved, both directly and as bystanders, might have incentives to alter accident related information to avoid punitive penalties. Therefore, it is critical to record authenticated event data at the specific time and location of the accident. The recorded information about the accident could later be retrieved and cross-examined to determine liability. It can provide the consensus on whether an event is verifiable and trustworthy at the certain geographic location and time. We propose the following two steps to accomplish the goal: first to gather trustable event data from both vehicles directly involved in the accident  and neighboring vehicles, then to verify and save event data with the help of a dynamically formed federation of vehicles within the same vehicular network.

\subsubsection{Gathering event data}
Vehicles directly involved in an accident are termed \say{accident} vehicles, vehicles within the DSRC transmission range from the accident scene are termed \say{witness} vehicles, and vehicles within the same cell but outside the DSRC transmission range from the accident scene are termed \say{community} vehicles. To record the event of an accident, upon the occurrence of an accident, \say{accident} vehicles send `event generation' requests to \say{witness} vehicles. Fig. \ref{fig:accident} depicts a scenario, where vehicles A and B got into an accident. \say{Witness} vehicles C, D, and E within the DSRC transmission range from the accident scene receive the event generation requests and confirm with the \say{accident} vehicles via DSRC. Fig. 4 depicts these sequence of events over time. 

\begin{figure}[h]
\includegraphics[width=0.45\textwidth]{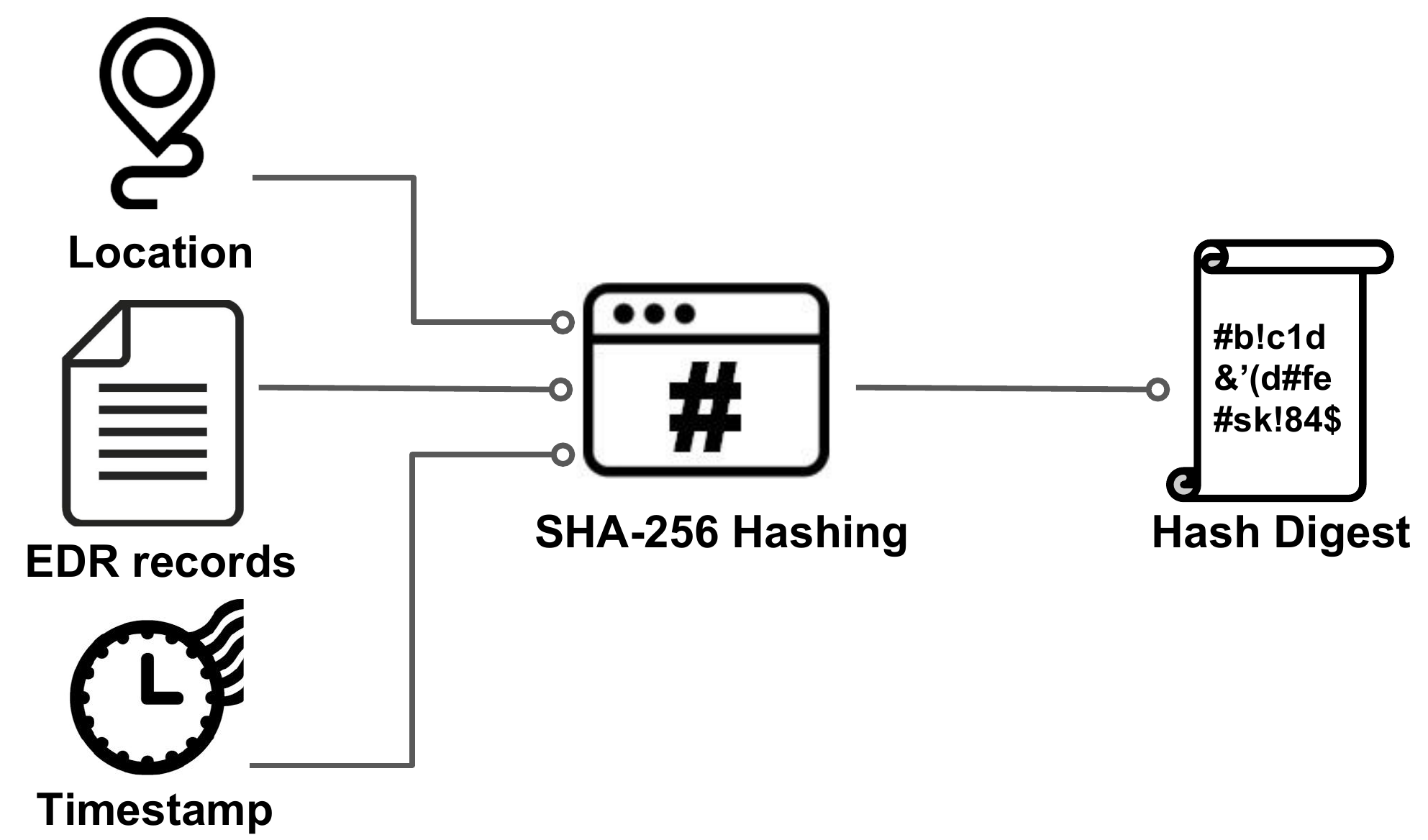}
\centering
\caption{Hash digest of event data.}
\end{figure}

Then, both \say{accident} and \say{witness} vehicles generate their respective event data with their own locations, EDR records (which capture histories of sensor readings around the accident scene), timestamps and the corresponding hash digests (as computed in Fig. 3), and broadcast their event data via cellular communications within the vehicular network under the same cell. All the broadcast event data from both \say{accident} and \say{witness} vehicles will be verified and saved in a new block by a federation to be described next. 

%While the \say{witness} cars received both requests from \say{accident} cars, A and B here, they understood that should provide the information about these two cars.
%The benefits of this mechanism are from two directions: individual concern and public concern. 
%\begin{figure}[h]
%\includegraphics[width=0.7\textwidth]{Figures/poenetwork}
%\centering
%\caption{Vehicle-to-vehicle proof of event network.}
%\end{figure}

%\begin{algorithm}
%\caption{{\it n} of {\it m} multi-signature federation consensus}\label{alg:FederationConsensus}
%\begin{algorithmic}[1]
%\Procedure{Federation}{$n,m$}\Comment{n of m model}
%\State $ ReceiveNewRecord$ 
%\While{$m>0$}
%\If{$HashDigestIsValid$}
%  \State $VerifyNewRecord(n)$
 % \If{$n/m > Threshold$}
 % \State $ SignNewBlock (n)$ \Comment{Event records confirmed %in a new block by n verifier vehicles}
 % \EndIf
 % \EndIf
%\State $AcceptNewBlock (m)$   
%\EndWhile\label{euclidendwhile}
%\State $BroadcastNewBlock(m)$ \Comment{New block has been broadcasted to network}
%\EndProcedure
%\end{algorithmic}
%\end{algorithm}

\begin{figure*}[h]
\includegraphics[width=0.88\textwidth]{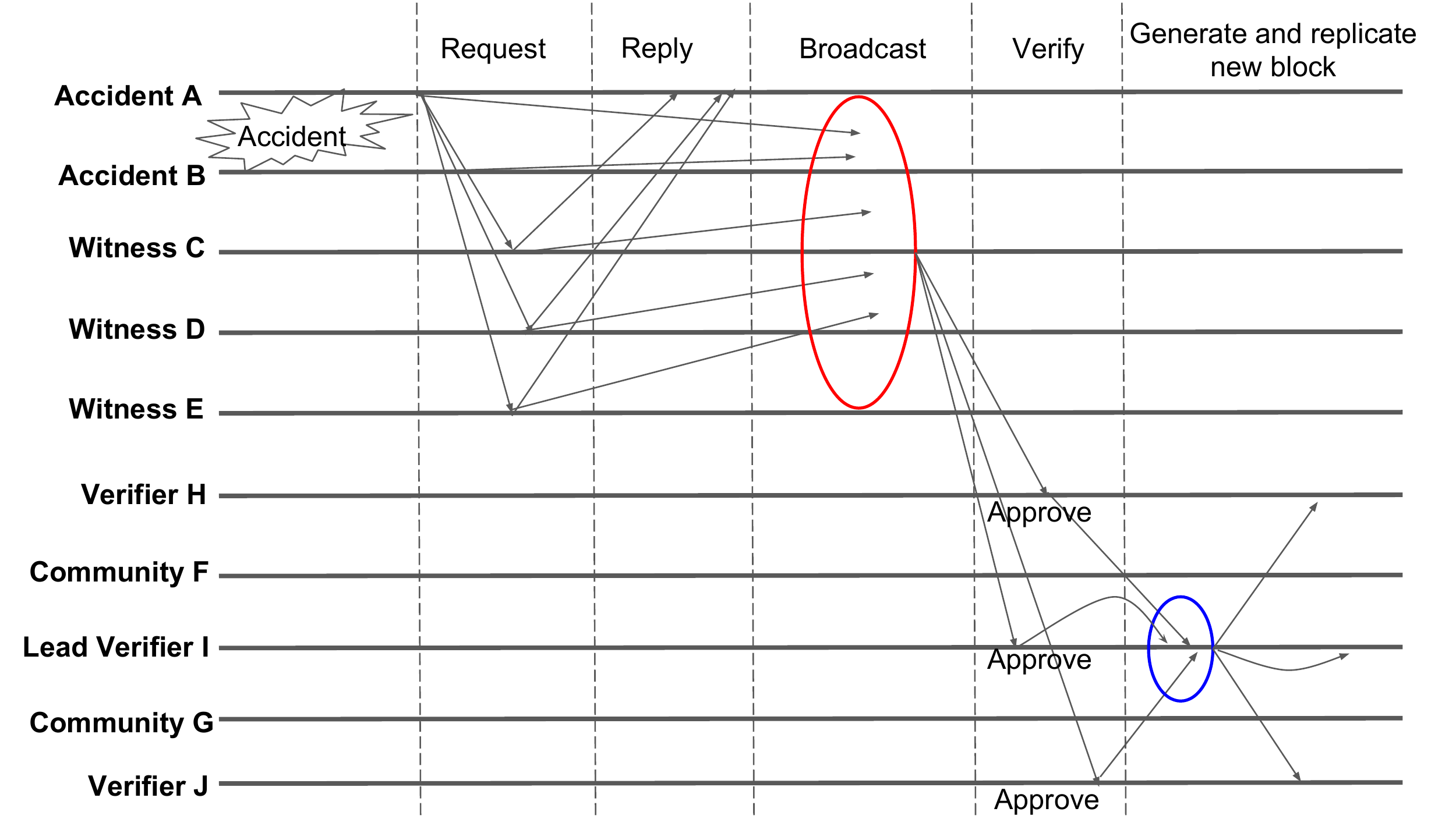}
\centering
\caption{Sequence diagram for accident, witness, community, and verifier vehicles.}
\label{fig:4}
\end{figure*}

\subsubsection{Verifying and creating new block of accident event}
Upon the occurrence of an accident, \say{accident} vehicles also broadcast, via cellular communication, `federation formation' requests to the \say{community} vehicles in the vehicular network to start the selection of a subset of \say{community} vehicles as \say{verifier} vehicles to form a federation. Such a selection process can be based on the notion of a {\em reputation score} which may be determined based on a vehicle's driving and reporting records. Also the \say{verifier} vehicle with the highest reputation score is designated as the {\em lead} verifier via a distributed leader election algorithm \cite{leader}, who is responsible for generating a new block for the accident. After that, new block generated by {\em lead} verifier vehicle will be sent to DMV and kept for the permanent records. 

As depicted in Fig. 4, after the \say{accident} and \say{witness} vehicles generate and broadcast event data into the vehicular network, \say{verifier} vehicles take the responsibility of validating the received event data against the received hash digests, and confirm with the lead \say{verifier} vehicle (I in Fig. 4). 
The lead \say{verifier} vehicle executes the {\it n}-of-{\it m} multi-signature scheme~\cite{bitcoinmulti} to achieve federation consensus when {\it n} out of {\it m} \say{verifier} vehicles confirm, and generates
% There are totally {\it m} \say{verifier} vehicles which will check the generated data with its hash digest to insure no one tamper the event data, and if at least {\it n} verifier from that federation group confirm the information, it can be accepted as 
a new block of accident event. The lead \say{verifier} vehicle may then broadcast the new block to all the \say{community} vehicles.

% However, increasing the threshold ({\it n} out of {\it m}) multi-signatures  provides stronger protection for the block, but reduces the flexibility of verifier vehicles in federation.
Compared to PoW, our solution does not incur any expensive computation associated with mining. 
Unlike PoA, which relies on the decision of one single authority, our solution demands confirmations from multiple authorities, if the {\it n}-of-{\it m} multi-signature~\cite{bitcoinmulti} threshold is satisfied, verifier vehicles approve the event records and generate a new block. Every new block has its hash header and linked with the previous block's hash header. For instance, as depicted in Fig. 5, block i is verified by 3 out of 4 verifier vehicles from federation, while the block i+1 is 3 out of 5 case.
% Comparing to PoA, which delivers transaction records through the consensus mechanism based on one single identity, our scheme avoids single authority crash or being malicious. Traditional Byzantine agreement~\cite{DBLP:conf/opodis/Lamport02} guarantees consensus. However, it requires agreement on identity by all peers. Every peer must be known and verified ahead of time. We achieve the byzantine robustness without central authority in a flexible way, and a minority of malicious participants will not be able to affect the system.
\begin{figure}[h]
\includegraphics[width=0.5\textwidth]{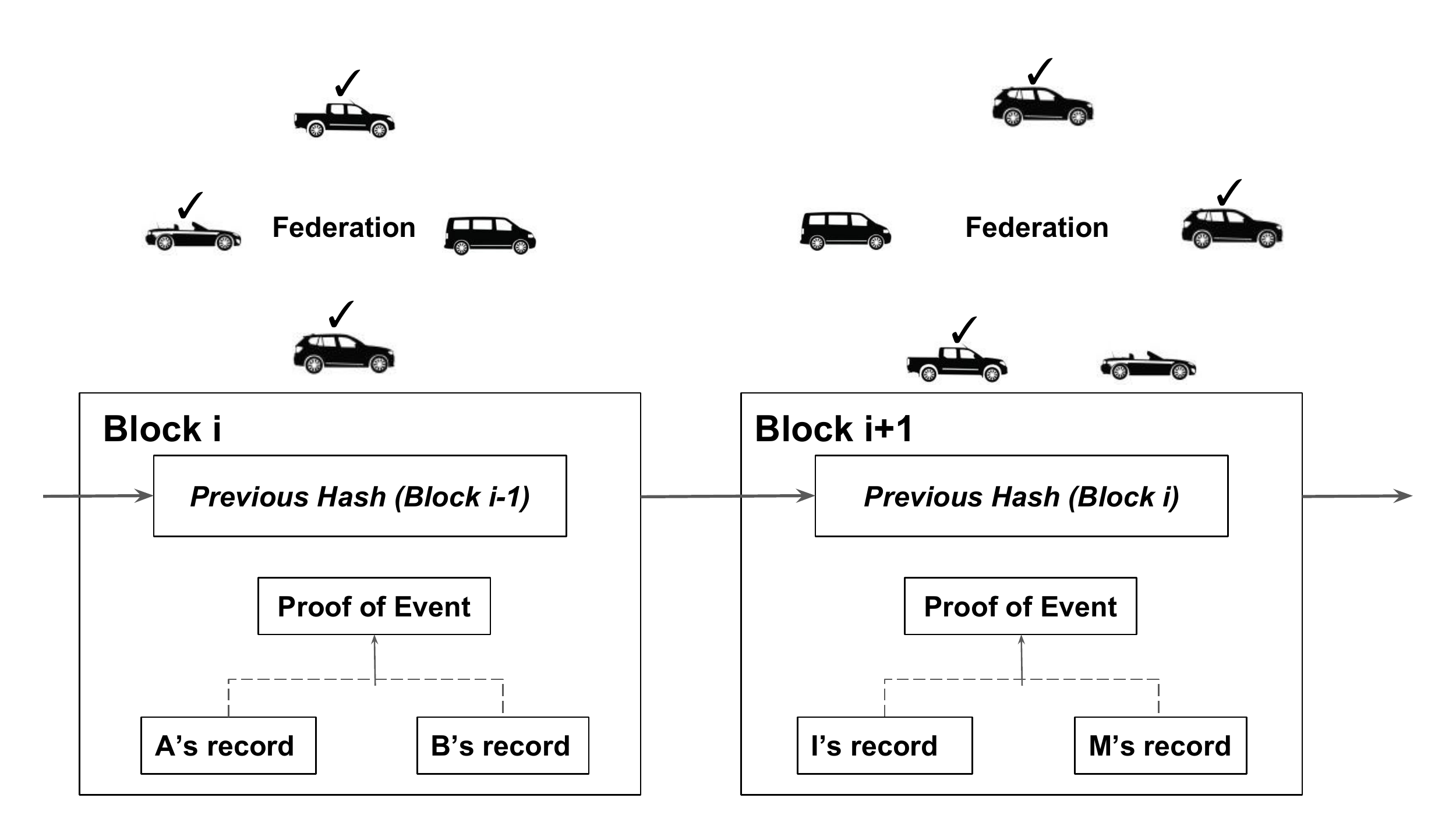}
\centering
\caption{Each block contains event data and hash value of previous block, and all confirmed by the verifier vehicles from the federation.}
\end{figure}

Note that \say{witness} vehicles (C, D, and E) function differently from \say{verifier} vehicles (H, I, and J). The job of the former is to generate event data, while that of the latter is to verify event data and generate a new block of accident event. \say{Witness} vehicles are close to the accident scene, whose EDR records may contain sensory readings related to the \say{accident} vehicles. In contrast, \say{verifier} vehicles are dynamically chosen which are located at random geographical locations within the same cell, even away from the accident scene, which makes them to be more neutral and independent. The decoupling of event data generations from their verification process mitigates the possibility of any malicious activities, such as tampering of event data and collusion among vehicles.

\subsection{Incentives for participation and honesty}

Bitcoin supplies new bitcoins to miners as an incentive for their efforts of PoW~\cite{tschorsch2016bitcoin}. However, there is no obvious tangible award in the proposed Proof of Event. To motivate autonomous vehicles to participate as either \say{witness} or \say{verifier}, different incentives (or rewards) must be defined. For instance, being a \say{witness} or \say{verifier}  could raise a vehicle's credit score and to lower its insurance premium. Also, \say{accident} vehicles that engage reliably in  Proof of Event and cooperate fully in accident forensics may receive a reduced liability.

\subsection{Reviewing blocks for accident forensics}

Later, people (police or judge) can review the accident event data stored in the blockchain from the DMV's record. If there is no discrepancy between event data generated by \say{accident} and \say{witness} vehicles, liability can be clearly determined. Otherwise, further investigation becomes necessary. For instance in Fig. 2, if \say{accident} vehicle B reported its own speed as 20 mph, while other \say{witness} vehicles (C, D, and E) reported higher speeds for B, it is highly likely that \say{accident} vehicle B had a faulty speed sensor which caused it to speed and collided with vehicle A.

%\begin{figure}[h]
%\includegraphics[width=0.4\textwidth]{Figures/blockverifiernew}
%\centering
%\caption{Federated  agreement 6-of-8 example.}
%\end{figure}

%Algorithms \ref{alg:Federated Consensus} presents how federation consensus works. Compared to traditional PoW scheme, our algorithm achieves significant reduction in the waiting time for confirming a new block. However, increasing the threshold provides stronger protection for the block, but reduces the resilience of verifier vehicles in the specific network.

%Here we highlight our proposal in several aspects:

%{\it Completeness}: The new block itself contains the correct previously hash digest, otherwise, it will be rejected immediately.
%{\it Number of Signatures}:
%The new block must contain the number of digital signatures generated from the verifier vehicles equal or greater than the threshold.

%{\it Integrity}:
%The set of block header hash value should not be the same as any header value before, even the new blocks are confirmed by the same federation of the verifier vehicles.

%{\it Timestamp}:
%The  new block timestamp must be greater than the previous block’s timestamp, otherwise it will be rejected immediately.
%\subsection{Reduce Penalty}
%In order to let every autonomous vehicle behave honestly, the proposed system should reduce the penalty and give possible incentives to an honest vehicle.

\section{Extreme Scenarios}
\label{sec:extreme}

Our proposed mechanism works the best in accident scenarios where the density of the cell-based vehicular network covering the accident scene is above a certain threshold. In such cases, there are enough \say{witness} to generate event data vehicles and enough \say{verifier} vehicles to form a federation, reach a consensus, and create a new block. However, there exists the following three `extreme' scenarios when such vehicular network is very sparse or no vehicle around the accident scene.  

{\em (1) Neither \say{witness} nor \say{verifier} vehicle exists for an accident}. Since there is no \say{witness} or \say{verifier} vehicle, no new block could be generated. The EDRs of the \say{accident} vehicles will be the only evidence for forensic investigation.

%Suppose one vehicle drives alone in the remote area and suddenly it hits an deer. Meanwhile, there is no other vehicles nearby when it sends the `event generation' request. So, it will only generate its own accident record and broadcast to vehicular network. In this case, we do not have any extra event information from other vehicles when check the event information in new generated block.

{\em (2) No \say{witness} vehicle exists for an accident}. A new block will be created based on the event data generated only by the \say{accident} vehicles.

{\em (3) No \say{verifier} vehicle exists for an accident}. In this case, there exists (few) \say{witness} vehicles around the accident scene, but no \say{verifier} vehicle within the vehicular network. We argue that such scenarios will be extremely rare in practice.  

\section{Security Analysis}
\label{sec:security}

In this section, we analyze the robustness 
of the proposed event recording scheme with respect to potential attacks.

\subsection{Spoofing event data}

Existing EDRs installed on individual vehicles may be hacked and tampered to avoid liability. Our proposed Proof of Event scheme have both \say{accident} and \say{witness} vehicles generate event data which is to be validated by an independent group of \say{verifier} vehicles to avoid possibility of collusion. The recored event data in the block could be cross-examined later to determine cause and liability.

Further, the use of DSRC communications range limits which vehicles could serve as witness, which prevents vehicles away from the accident scene to generate any `fake' event data.

\subsection{Impersonation attack}

As mentioned in Section \ref{sec:cell}, legitimate autonomous vehicles are required to register with DMV. A malicious vehicle may impersonate a reputable vehicle so as to be selected as a \say{verifier} vehicle. Unless there are enough number of colluding vehicles selected within the same federation, the use of {\it n}-of-{\it m} multi-signature scheme to approve a new block is to lower the possibility of invalidate consensus.

\subsection{Fake witness vehicle attack}

As mentioned before, `event generation' requests are broadcast via DSRC so that only the \say{witness} vehicles, which receive such request, can generate  event data. However, a \say{witness} vehicle might resend the `event generation' request to other vehicles which are beyond the range of DSRC communication, and `invite' them to respond. Such act may launch the fake \say{witness} vehicle attack, where fake \say{witness} vehicles generate the fake event data in favor of \say{accident} vehicles. 
One possible solution to prevent such attacks is to set a small time window for \say{witness} vehicles to reply and broadcast event data.

% or each \say{witness} vehicle should state and send the ID of other legitimate \say{witness} vehicles who are within the valid range from the accident scene. 

\section{Conclusion}
\label{sec:conclusion}

% With autonomy, comes accountability. 

As autonomous systems are becoming essential parts of our life, proper systems must be put in place to ``look after'' them so as to determine liability from malfunctions, defects, or even malicious attacks.
By drawing inspiration from blockchain, this paper presents a novel approach to providing a tamper-proof and verifiable event recording system for accident forensics of self-driving vehicles as they are the most influential autonomous systems in our society. In future, we plan to evaluate our proposed scheme using an experimental testbed.

% We illustrate our proposal with several benefits which guarantees both robustness and efficiency with low power consumption. Furthermore, we conduct the security analysis against potential attacks.
% In future work, we plan to enable vehicle-to-X communication by connecting the vehicles with road-side-unit (RSUs) and other IoT devices, to record the information in a comprehensive way.
%we plan to implement the proposed scheme in a large vehicular network and evaluate the performance. 
 


%\bibliographystyle{IEEEtran}

\bibliographystyle{IEEEtran}
\bibliography{bibtex/IEEEabrv,bibtex/sig}

%\begin{thebibliography}{1}


%\end{thebibliography}

\end{document}